Horváth István,[1]

# VERIFYING GAMMA-RAY BURST DURATION ANALYSIS USING THE COMPTON GAMMA-RAY OBSERVATORY T50 DATA [2]

*Several authors have shown in the literature that a third type of gamma-ray burst class can be detected in the Compton Gamma-Ray Observatory data using T90. In this paper, I examine whether this can also be found in the T50 variable data from the Compton Gamma-Ray Observatory Burst And Transient Source Experiment observation. According to my calculations, the third group detected in T90 can also be detected in the T50 data.*
***Keywords:*** *gamma-ray burst, gamma-ray, satellites, statistical analysis*

## Introduction

Since the discovery of gamma-ray bursts in the last century, many satellites have been equipped with gamma-ray detectors to observe the bursts. One of the satellites built for this purpose is the Compton Gamma-Ray Observatory. It has several instruments on board, one of which is the Burst And Transient Source Experiment. The Burst And Transient Source Experiment has observed more than 2,000 gamma-ray bursts during its nearly ten-year operation. The individual flashes can be divided into two groups according to their duration[3]. One type is a short burst, the other type is bursts lasting longer than two seconds.

## Determining the duration of gamma-ray bursts

Defining the duration of gamma-ray bursts is not an easy task, as we cannot know from which direction each photon arrived, because the Burst And Transient Source Experiment detectors could not measure direction, meaning that the background photons and the photons arriving from the source are indistinguishable. We count the photon arrivals and then classify them into time slots, i.e. we count the number of incoming photons during a chosen time unit. This gives the high-energy temporal distribution of the gamma-ray burst.

This procedure allows us to measure time as a multiple of the aperture length. If a data series with better time resolution is available, then the cumulative curve can of course be generated with finer resolution.

The beginning of a gamma-ray burst is usually identifiable, but the end of the burst is more difficult to determine. The initial and final levels of the integral light curve are better defined. If the background is still well-fitted after the burst, then the right-hand side of the integral curve will also fluctuate around a definite value. This value is the total number of photons from the burst that the detector has recorded. By intersecting the integral curve with one twentieth of this number, we get the time when the detector has registered 5 percent of the total photons from the burst. Similarly, we can determine the time when 95 percent of the total photons have arrived. The difference between these two times is called T90. The value of T50 is naturally determined by the times corresponding to 25% and 75%. In 1995, the figure published by Kouveliotou et al. shows two Gaussian curve fits (Figure 1).[4]

[1] Professor, NKE HHK Természettudományi tanszék, e-mail: horvath.istvan@uni-nke.hu ORCID: https://orcid.org/0000-0002-1343-1761

[3] NORRIS 1984. KOUVELIOTOU 1993.

[4] KOUVELIOTOU 1995.

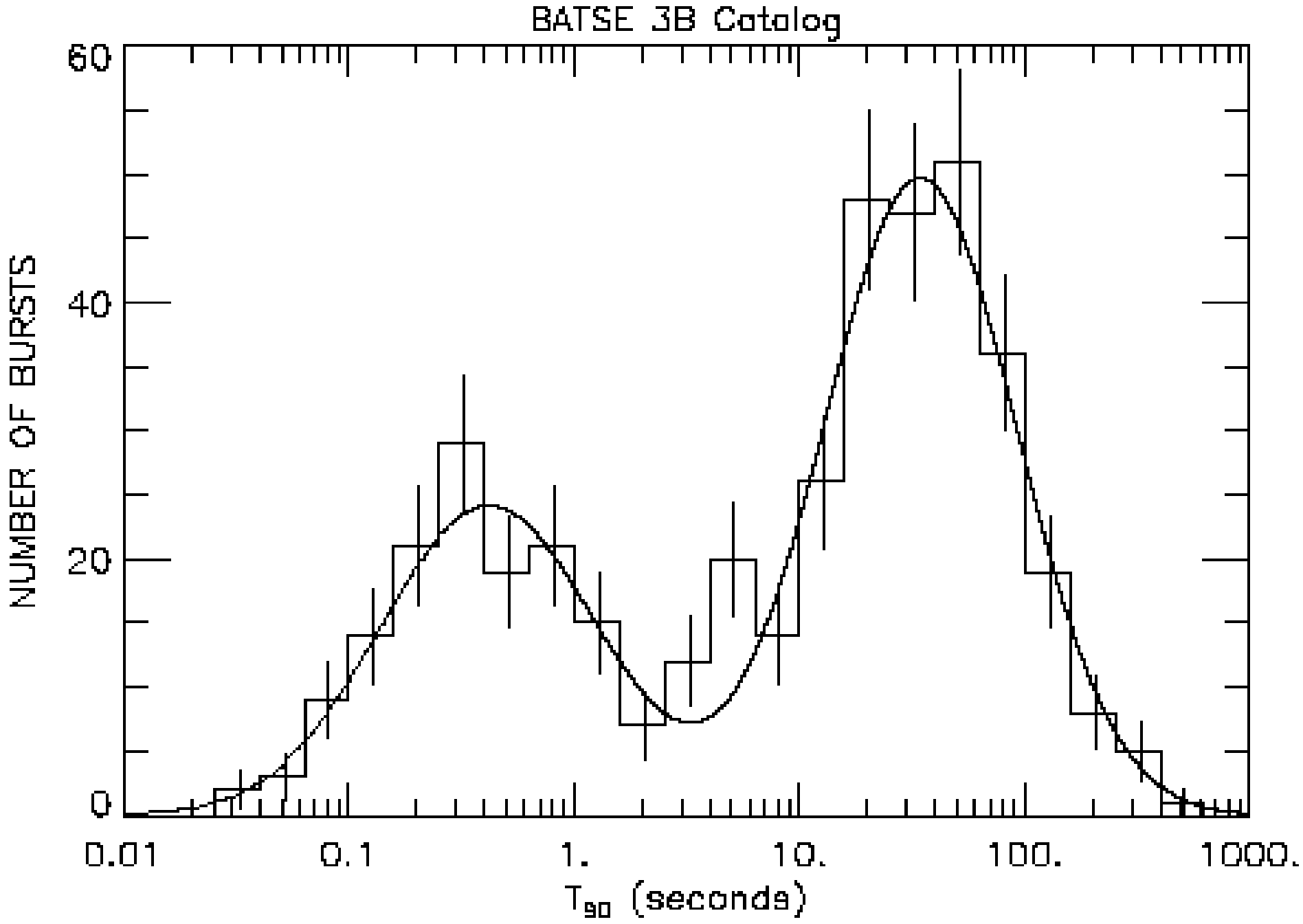

*1. figure. Duration frequency of bursts in the Burst And Transient Source Experiment 3B catalog.*
*Source:* KOUVELIOTOU 1995.

The third Burst And Transient Source Experiment catalog was published in 1996, and the data became available online almost immediately. The Burst And Transient Source Experiment catalogs are made up of several sections, three of which have detailed tables that are still available online today.[5].
The “basic” table contains basic information, such as the catalog number or the date and exact time of the observation. The “Flux and Fluence” table contains the identification number as well as the fluence value, which shows how much energy was received in the given frequency range during the entire outburst. This is essentially the integration of the light curve. The peak flux, in Hungarian the peak brightness, indicates the highest value of the light curve.
The Burst And Transient Source Experiment 3B catalog contained a total of 1122 gamma-ray bursts. The duration table included data for 834 bursts.
I analyzed these in a previously published article[6]. The duration distribution shows several peaks. There is a distinct group of long and short bursts. I also examined the existence of additional classes in the aforementioned article. During the analysis I used $\chi^2$

$$\chi^2 = \sum_{i=1}^{N} \frac{\left(f_i - f\left(x_i\right)\right)^2}{\sigma_i^2} \qquad (1)$$

where $f_i$, is the fitted function, $f(x_i)$ the measurement results and $\sigma_i$ is a standard deviation.
If the distribution of a variable spans several orders of magnitude, the logarithm of the variable is usually used in the fitting. The logarithmic distribution of gamma-ray burst durations can be modeled using Gaussian curves. To determine whether an additional component was needed, I used the following method. I fitted a theoretical curve to the entire distribution, which consists of the sum of three Gaussian curves (Figure 2). The resulting fit had a $\chi^2$ value of 24.0. This result should be compared with the best $\chi^2$ value (46.8) from the previous fit using the sum of two Gaussian curves.
The $\chi^2$ difference between the two fits is 46.8 - 24.0 = 22.8, which determines the probability. Since the model with three Gaussians introduces three additional parameters, the probability associated with this follows a $\chi^2$ distribution with three degrees of freedom. In this case, the probability associated with 22.8 is extremely low, only $10^{-4}$. This means that the observed distribution has an extremely low chance of occurring by chance.

[5] https://gammaray.msfc.nasa.gov/batse/grb/catalog/ (2025. 03. 01.)
[6] HORVÁTH 1998.

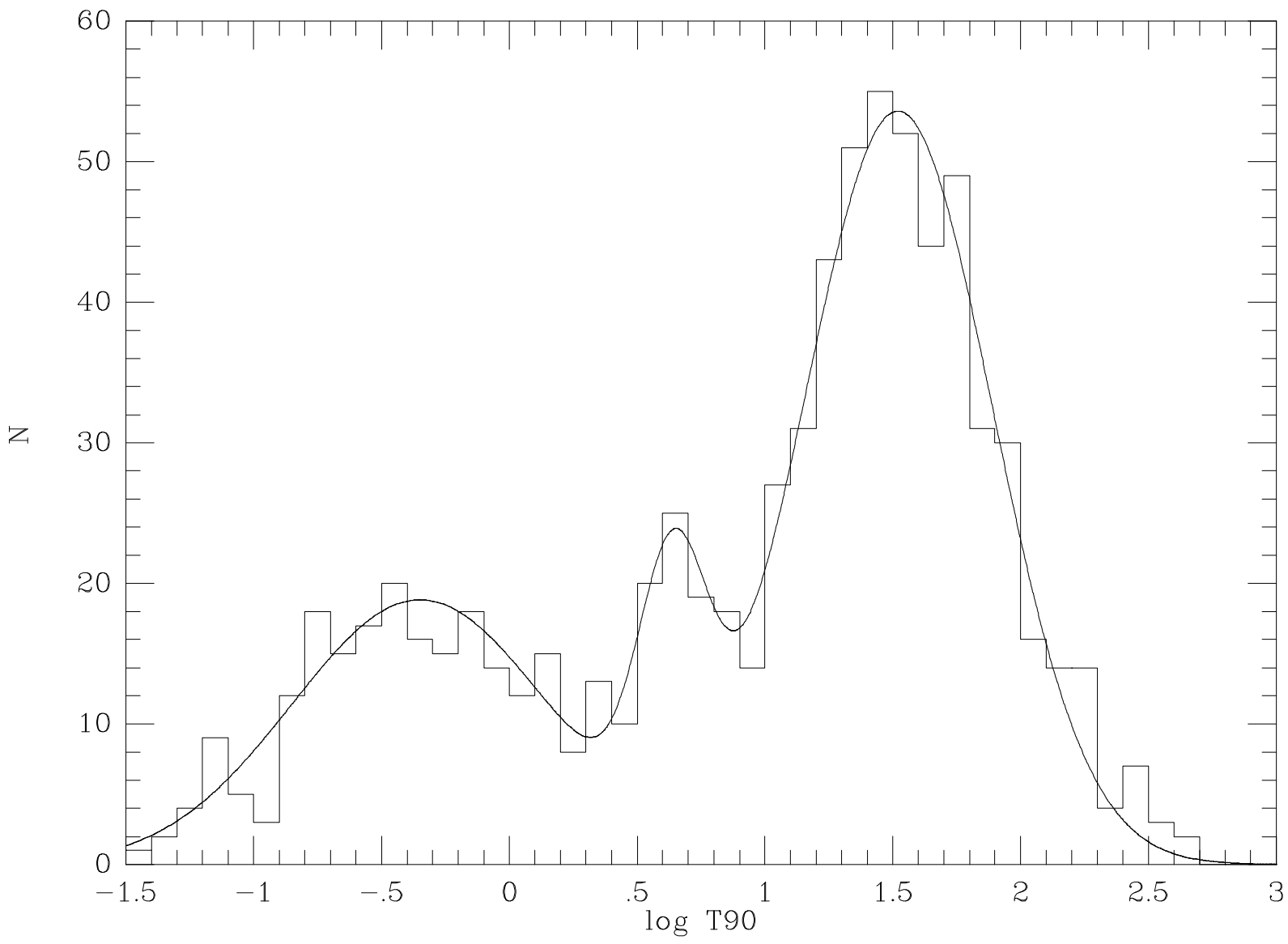

*2. figure. The three-component fit to the T90 duration distribution*.
*Source:* HORVÁTH 1998

Table 1 contains the parameters of the best fit. Based on the data, the average T90 duration of the so-called long gamma-ray bursts is approximately 33 seconds. According to the fit results, 62% of the observed gamma-ray bursts fall into this category. For short bursts, the average T90 duration is about half a second, and about one-third of the gamma-ray bursts detected by the Burst And Transient Source Experiment can be classified into this group. Based on the fits to the duration distribution, a significant proportion of gamma-ray bursts between 2 and 8 seconds can be classified into a third category. Since these events are longer than 1-2 seconds but do not reach the 10-50 second time range, it seems justified to use the term intermediate-type gamma-ray bursts.

**1 Table.** Parameters of the three groups.

| | Average value (lg T90) | standard deviation | no. of GRBs | % | typical T90 |
|---|---|---|---|---|---|
| short | -0,35 | 0,50 | 236 | 30 | < 2 mp |
| long | 1,52 | 0,37 | 497 | 62 | > 8 mp |
| interm. | 0,64 | 0,14 | 61 | 8 | ~ 2-8 mp |

**Analysis of the data from the definitive Burst And Transient Source Experiment catalog**
The Compton satellite ended its operation in 2000. The Burst And Transient Source Experiment recorded 2704 gamma-ray bursts. The observed data can be accessed on the Internet in the Burst And Transient Source Experiment Final Catalog1. The duration table contains data for 1929 bursts. In 2002, I examined the data for these 1929 bursts2. For this analysis, I used the maximum likelihood method. Thus, each data point is included in the analysis by itself. The essence of the maximum likelihood method is that the true value of the parameter a is estimated by the value a' that, if the parameter had a true value, would make the occurrence of the given pattern the most likely among all possible patterns.
If we are given an n-element measurement sample x1, x2, …. xn for a variable whose density function is

$$f = f(x, a) \qquad (2)$$

and we want to estimate the value a from the measurement results, then the

$$l(x_1, x_2, \ldots x_n, a) = f(x_1, a) f(x_2, a) \ldots f(x_n, a) \qquad (3)$$

function is called the likelihood function.
The maximum likelihood method is to choose the value where the value of the likelihood function is maximum to estimate the parameter. Among the asymptotically normal distribution estimates, the maximum likelihood estimate is the best estimate[7].
In the analysis of the 1929 data from the Burst And Transient Source Experiment final catalog mentioned above, a Gaussian curve fit, similarly to the previous ones, did not give a match. When fitting two Gaussian curves, there were 5 parameters to be fitted. A Gaussian curve can be described by 3 parameters; the maximum location, the standard deviation and the weight of the distribution. Consequently, in the case of two Gaussian curves, we have 6 parameters to be fitted, but there is also a condition for the sum of the amplitudes. The sum of the amplitudes is either N (the number of detected objects), or if the detection probability is the fitted function, then the value of the total integral is 1. So the aforementioned condition can be satisfied if the sum of the weight coefficients used during the fit is 1.
In the case of fitting two Gaussian curves, the maximum value of L is 12320.11, and in the case of fitting three Gaussian curves, the maximum value of L is 12326.25. An improvement in the maximum value of the likelihood is expected, since we describe the fitted theoretical curve with more parameters. In this case, the number of parameters has increased by 3. In this case, twice the difference in the L values follows a χ2 distribution with 3 degrees of freedom[8].
The probability calculated based on this

$$2(L_3 - L_2) = \chi_3^2 \qquad (4)$$

formula is 0.5%.

[7] BRONSTEIN, I. N. AND SZEMENGYAJEV I 1980. PAÁL 1992.
[8] KENDALL, M. AND STUART 1976.

The parameters obtained from the fits are shown in Table 2 and Table 3. The significance was checked using the Monte-Carlo method.

**2 Table.** Parameters of short and long groups.

| group | lgT90 center | standard deviation | % |
|---|---|---|---|
| short | -0,11 | 0,61 | 0,32 |
| long | 1,54 | 0,43 | 0,68 |

**3 Table.** Parameters of the three groups.

| group | lgT90 center | standard deviation | % |
|---|---|---|---|
| short | -0,25 | 0,53 | 0,26 |
| interm. | 0,63 | 0,20 | 0,06 |
| long | 1,55 | 0,42 | 0,68 |

Using the data in Table 2 as a density function, I randomly selected 1929 numbers. I performed the two-component and three-component fitting on these numbers. Of course, the likelihood value was higher in the case of three components. I performed this procedure 99 more times (with 99x1929 random numbers). The difference distribution of the hundred likelihoods is shown in Figure 3. During the procedure, I assumed that I would select 1929 points from a two-component sample. Assuming that there is no third component, I use the Monte-Carlo procedure to examine the probability of obtaining the improvement in likelihood experienced in the measurement (6.14). This happened once in a hundred cases, which confirms the 0.5% probability obtained using the $\chi 2$ distribution with 3 degrees of freedom.
That the third population is not the result of random fluctuation is also supported by the fact that the group size was larger than the six percent obtained during the measurement only once out of a hundred cases. The average group size during the Monte-Carlo simulation was 2.5%.

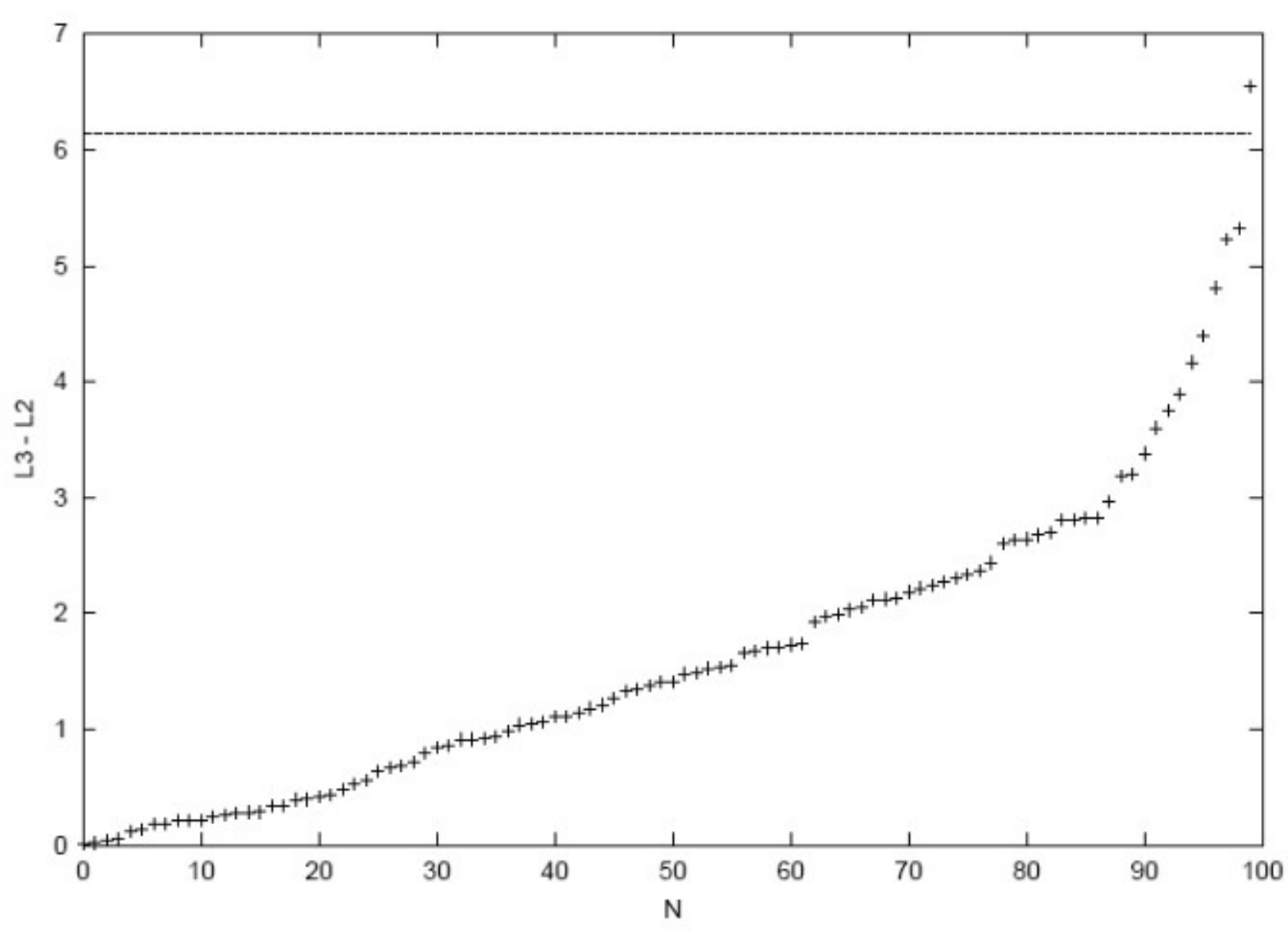

*3. figure. Distribution of Monte-Carlo simulated likelihood differences sorted in ascending order. The value of 6.14 indicated by the dashed line is the likelihood improvement obtained in the analysis. Only one of the Monte-Carlo simulations exceeded this value.*
*Source:* HORVÁTH 2002

**Studies using T50 data**

First, I repeated the calculations using the maximum likelihood method on the T90 data of the 1929 outburst of the complete Burst And Transient Source Experiment catalog. The result was completely consistent with the results reported in the literature. Assuming two groups, the maximum likelihood was 12320.11, and assuming three Gaussian components, it was 12326.25. The difference was 6.14, which corresponds to a significance of 99.5%. The best-fitting parameters were also identical to those previously reported in the literature. I also performed the calculations on the 1929 T50 data. Assuming two groups, the maximum likelihood was 12216.14, and assuming three Gaussian components, it was 12221.61. The difference was 5.47, which corresponds to a significance of 98.8%.
In 1998, I demonstrated the third component in the T90 distribution of 797 bursts in the Burst And Transient Source Experiment 3B catalog with 99.98% significance1. I note here that using data from the same 797 gamma-ray bursts, another group demonstrated the existence of the third component at the same time[9]. In their detailed fourteen-page article, in addition to the two hardness factors, peak brightness and total flux, they used not only T90 but also T50, not separately, but all six variables at once. They found three groups in this six-dimensional space. The projection of their results onto the time axis showed a high degree of agreement with the results I had previously published.

I have now re-fitted the distribution of the T90 and T50 data of the 797 gamma-ray burst. The fitting was performed using the maximum likelihood method. In the case of T90, assuming two groups, the maximum likelihood was 4364.52, and assuming three Gaussian components, it was 4375.12. The difference is 10.6, which corresponds to a significance of 99.99%. The group parameters are contained in Table 4.

**4 Table.** For three components and T90, the resulting group parameters are: ql is the group weight ratio.

| csoport | $q_l$ | *log T90* | *standard deviation* |
|---|---|---|---|
| short | 0,28 | -0,35 | 0,49 |
| interm. | 0,07 | 0,61 | 0,14 |
| long | 0,65 | 1,53 | 0,40 |

In the case of T50, the maximum likelihood was 4341.6 assuming two groups, and 4354.5 assuming three Gaussian components. The difference is 12.9, which corresponds to a significance of 99.999%. Thus, the T50 values show the existence of the third group with even greater significance. The resulting group parameters are contained in Table 5.

**5 Table.** For three components and T50, the group parameters are: ql is the fraction of the group in all observed gamma-ray bursts.

| csoport | $q_l$ | *log T50* | *standard deviation* |
|---|---|---|---|
| short | 0,27 | -0,76 | 0,45 |
| interm. | 0,05 | 0,22 | 0,09 |
| long | 0,68 | 1,05 | 0,50 |

Comparing the two tables, we can see that the two analyses give essentially the same result. By doing so, I have now shown that the Burst And Transient Source Experiment T50 data also confirm the existence of the third gamma-ray burst group.

[9] MUKHERJEE 1998.

**Summary**
The precise knowledge of the sources of gamma-ray bursts is aided by their classification. It is widely believed that there are two types of sources for the bursts. Long bursts may be associated with the final states of massive, rapidly rotating stars, while short bursts may be produced by the merger of compact binaries. Several attempts have been made to identify a type of gamma-ray burst that is different from the two known types (e.g. Pendleton et al., 1997.1). There is an extensive literature on the possible group of intermediate-duration gamma-ray bursts.[10]
In my previous work, I examined the distribution of the durations of 797 gamma-ray bursts in the Burst And Transient Source Experiment 3B catalog. I showed that the three-component distribution is a significant improvement over the two-component one. I then analyzed the final Burst And Transient Source Experiment catalog. I examined the duration distribution of 1929 gamma-ray bursts in the catalog and showed that the three-component distribution is a significant improvement over the two-component one.
In this article, I analyzed the T50 data from both the 3B and the definitive Burst And Transient Source Experiment catalogs. Similar to the T90 studies, I was able to show that the T50 data also confirm the existence of the third gamma-ray burst group.

[10] BALASTEGUI, A. – RUIZ-LAPUENTE, P. AND CANAL 2001. HORVÁTH 2002. MÉSZÁROS 2006. HORVÁTH 2009. ZITOUNI 2015. TARNOPOLSKI 2016. HORVÁTH 2020.

## A GAMMAKITÖRÉS HOSSZÚSÁG ADATOK VIZSGÁLATAINAK ELLENŐRZÉSE A CGRO T50 IDŐTARTAMOK HASZNÁLATÁVAL

*A szakirodalomban többen megmutatták, hogy a Compton Gamma-Ray Observatory adataiban kimutatható egy harmadik típusú gammakitörés osztály a T90 változót használva. Jelen cikkben megvizsgálom, hogy a* Compton Gamma-Ray Observatory Burst And Transient Source Experiment *mérésiadataiban szereplő T50 változó adataiban is fellelhető-e ez a tendencia. Számításaim szerint a T90-ben kimutatott harmadik csoport kimutatható a T50 adatokban is.*

**Bevezetés**

A gammakitörések múlt századi felfedezése óta sok műholdon helyeztek el gammadetektorokat, a kitörések megfigyelésére. Az egyik erre a célra készített műhold a Compton Gamma-Ray Observatory. Fedélzetén több műszer kapott helyet, melyek egyike a Burst And Transient Source Experiment volt. A Burst And Transient Source Experiment műszer közel tíz éves működése során több mint 2000 gammakitörést figyelt meg. Időtartamuk szerint az egyes felvillanások két csoportba sorolhatók[11]. Az egyik típus a rövid kitörés, a másik típus a két másodpercnél is tovább tartó kitörések.

**A gammakitörések időtartamának meghatározása**

A gammakitörések időtartamának a definiálása nem könnyű feladat, ugyanis az egyes fotonokról nem tudhatjuk, hogy milyen irányból érkeztek, mert a Burst And Transient Source Experiment detektorok nem tudtak irányt mérni, vagyis a háttérfotonok és a forrásból érkező fotonok megkülönböztethetetlenek. A foton beérkezéseket leszámoljuk, majd időrekeszekbe soroljuk ezeket, azaz egy választott időegység alatt megszámoljuk a beérkező fotonok számát. Ez adja a gammakitörés nagy energiás időbeli eloszlását.

Ebből az eljárásból adódóan az időt a rekeszhossz többszöröseként tudjuk mérni. Ha rendelkezésre áll jobb időfelbontású adatsor, akkor természetesen a kumulatív görbe előállítható finomabb felbontással is.

Gammakitörés kezdete általában azonosítható, de a kitörés végét nehezebb meghatározni. Az integrális fénygörbe kezdeti és végső szintje már jobban megadható. Ha a háttér a kitörés után is jól illeszthető, akkor az integrális görbe jobb oldala is egy határozott érték körül ingadozik. Ez az érték a kitörésből származó összes fotonok száma, melyet a detektor rögzített.

E számnak a huszadával elmetszve az integrális görbét, megkapjuk azt az időpontot, amikor a detektor az összes, a kitörésből érkező foton 5 százalékát regisztrálta. Hasonlóan meghatározható, hogy mikor volt az az időpont, amikor az összfotonszám 95 százaléka beérkezett. E két idő különbségét nevezik T90-nek. A T50 értékét értelemszerűen a 25% és 75%-nak megfelelő idők határozzák meg. 1995-ben, Kouveliotou és társai[12] által közölt ábrán két Gauss-görbés illesztés látszik (1. ábra).

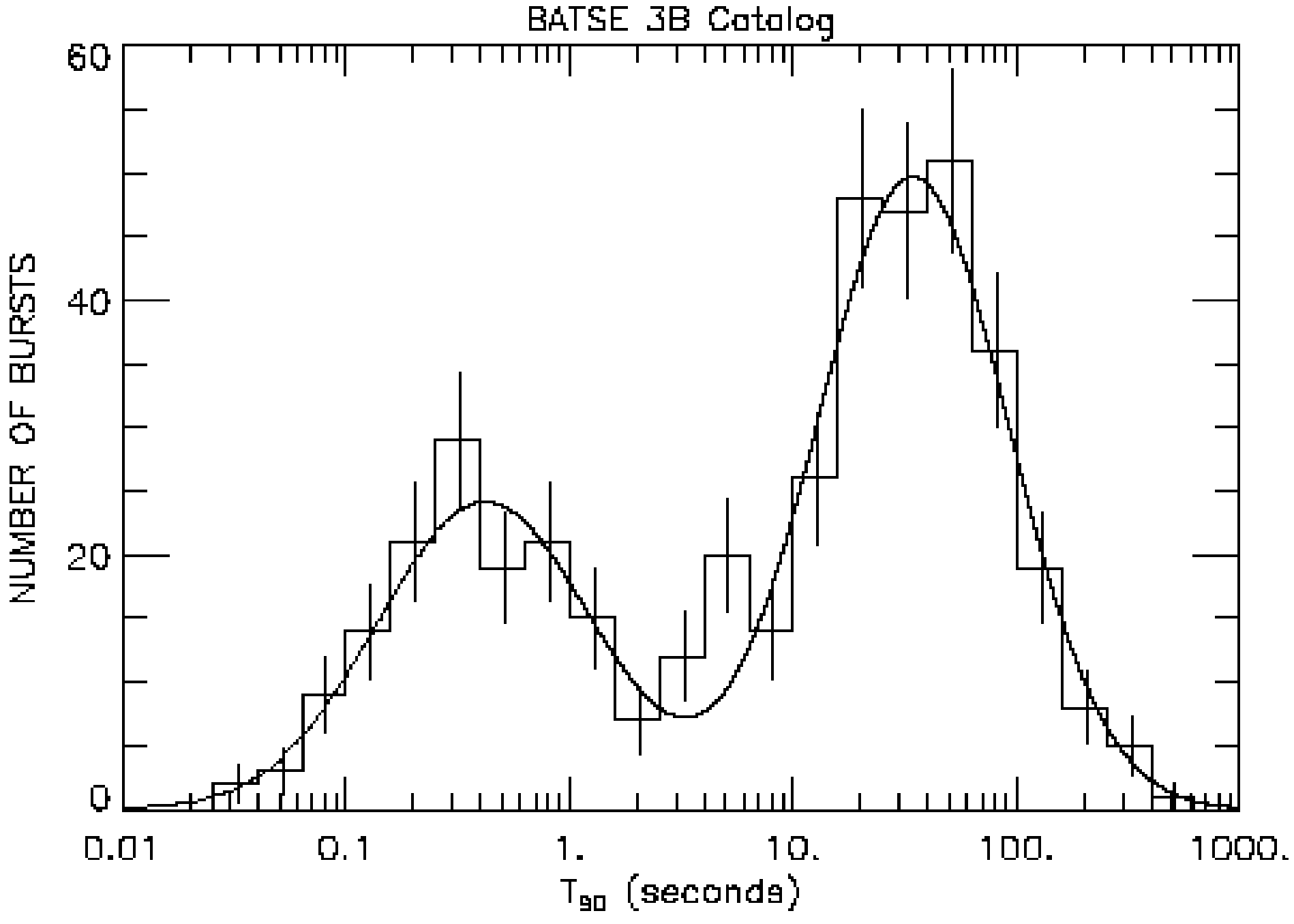

*1. ábra.* A Burst And Transient Source Experiment 3B katalógus kitöréseinek időtartamgyakorisága.

[11] NORRIS 1984. KOUVELIOTOU 1993.

[12] KOUVELIOTOU 1995.

*Forrás:* KOUVELIOTOU 1995.

A harmadik Burst And Transient Source Experiment katalógust 1996-ban tették közzé[13], és az adatok szinte azonnal elérhetővé váltak az interneten. A Burst And Transient Source Experiment katalógusok több szekcióból épülnek fel, amelyek közül három részletes táblázat az interneten azóta is elérhető[14].
A „basic” táblázat alapvető információkat tartalmaz, például a katalógusszámot vagy a megfigyelés dátumát és pontos idejét. A „Flux and Fluence” táblázatban az azonosítószám mellett a fluenciaérték is szerepel, amely azt mutatja meg, hogy a teljes kitörés során mekkora energia érkezett be az adott frekvenciatartományban. Ez lényegében a fénygörbe integrálja.
A peak flux, magyarul csúcsfényesség a fénygörbe legnagyobb értékét jelöli.
A Burst And Transient Source Experiment 3B katalógus összesen 1122 gamma-kitörést tartalmazott. Az időtartam-táblázat 834 kitörés adatait foglalta magában.
Ezeket elemeztem egy régebben publikált cikkemben[15]. Az időtartam eloszlás több csúcsot is mutat. Elkülönül a hosszú és a rövid kitörések csoportja. További osztályok létezését is megvizsgáltam az említett cikkben. A vizsgálatok során a $\chi^2$-próbát alkalmaztam, ahol a $\chi^2$ következő módon számolandó

$$\chi^2 = \sum_{i=1}^{N} \frac{\left(f_i - f\left(x_i\right)\right)^2}{\sigma_i^2} \qquad (1)$$

itt $f_i$ az illesztett függvény, $f(x_i)$ a mérési eredmények és $\sigma_i$ a hiba.
Ha egy változó eloszlása több nagyságrendet lefed, akkor rendszerint a változó logaritmusát használjuk az illesztés során. A gammakitörések időtartamának logaritmikus eloszlását Gauss-görbékkel modellezhetjük. Annak eldöntésére, hogy szükség van-e egy további komponens bevezetésére, a következő módszert alkalmaztam. A teljes eloszlásra egy olyan elméleti görbét illesztettem, amely három Gauss-görbe összegéből áll (2. ábra). Az így kapott illesztés $\chi^2$ értéke 24,0 lett. Ezt az eredményt a korábban, két Gauss-görbe összegével végzett illesztés legjobb $\chi^2$ értékével (46,8) kell összehasonlítani.
A két illesztés közötti $\chi^2$ különbség 46,8 - 24,0 = 22,8, meghatározza a valószínűséget. Mivel a három Gauss-görbét tartalmazó modell három további paramétert is bevezet, az ehhez tartozó valószínűség egy három szabadsági fokú $\chi^2$-eloszlást követ. Ebben az esetben a 22,8-hoz tartozó valószínűség rendkívül alacsony, mindössze $10^{-4}$. Ez azt jelenti, hogy a megfigyelt eloszlás véletlenszerű kialakulásának esélye rendkívül csekély.

---

[13] MEEGAN 1996.
[14] https://gammaray.msfc.nasa.gov/batse/grb/catalog/ (Letöltve: 2025. 03. 01.)
[15] HORVÁTH 1998.

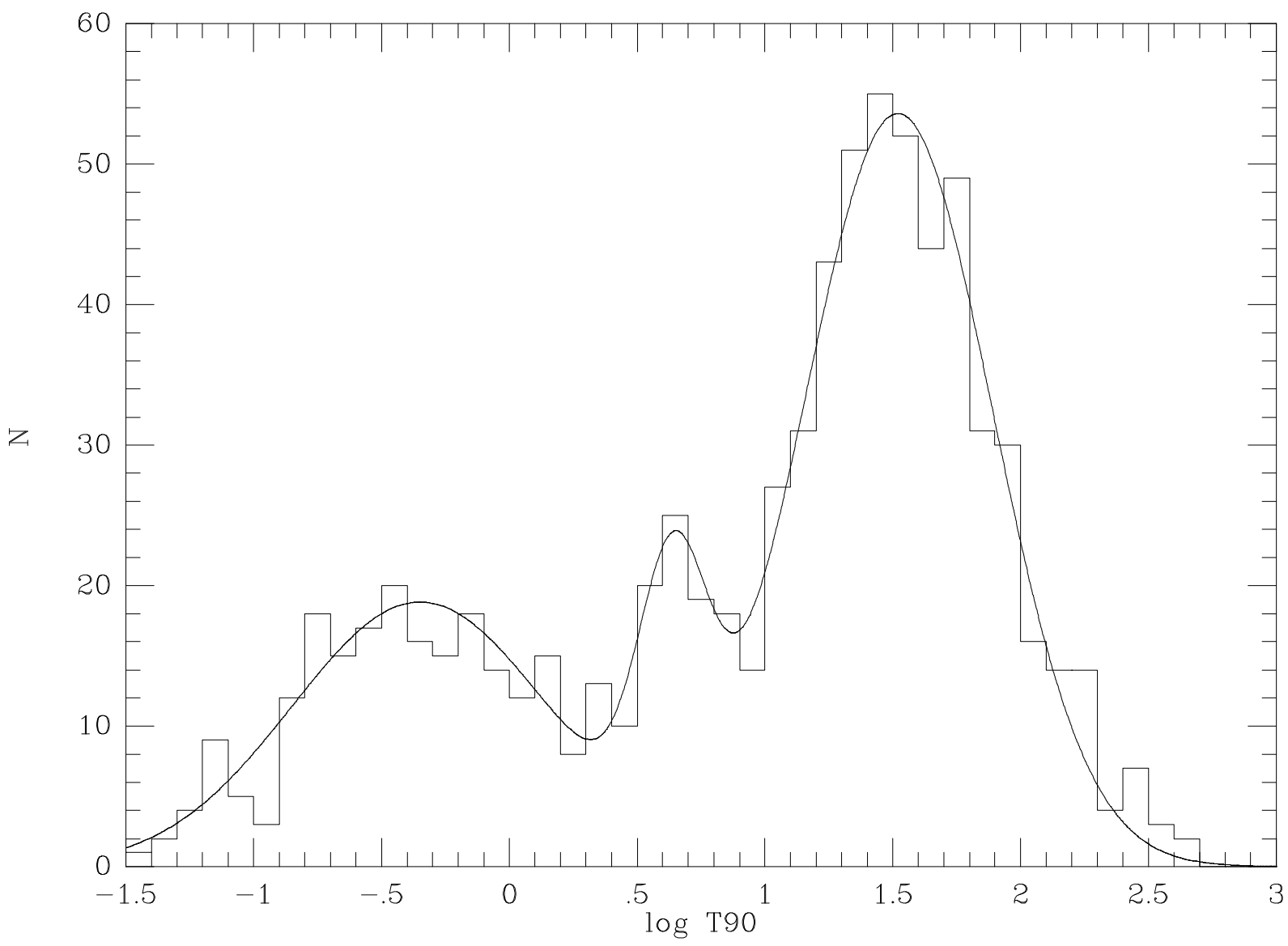

*2. ábra.* A háromkomponensű illesztés a T90 időtartameloszlásra.
*Forrás:* HORVÁTH 1998

Az 1 táblázat tartalmazza a legjobb illesztés paramétereit. Az adatok alapján az úgynevezett hosszú gammakitörések átlagos T90 időtartama megközelítőleg 33 másodperc. Az illesztési eredmények szerint a megfigyelt gammakitörések 62%-a ebbe a kategóriába tartozik.
A rövid kitörések esetében az átlagos T90 időtartam körülbelül fél másodperc, és a Burst And Transient Source Experiment által detektált gammakitörések nagyjából egyharmada sorolható ebbe a csoportba. Az időtartamok eloszlására végzett illesztések alapján a 2-8 másodperc közötti gammakitörések jelentős része egy harmadik kategóriába eshet. Mivel ezek az események hosszabbak, mint 1-2 másodperc, de nem érik el a 10-50 másodperces időtartományt, indokoltnak tűnik a közepes típusú gammakitörések megnevezés alkalmazása.

**1 táblázat.** A három csoport paraméterei.

| | Középérték (lg T90) | a lgT90 szórása | tagok száma | részarány %-ban | tipikus időtartam T90-ben |
|---|---|---|---|---|---|
| rövid | -0,35 | 0,50 | 236 | 30 | < 2 mp |
| hosszú | 1,52 | 0,37 | 497 | 62 | > 8 mp |
| közepes | 0,64 | 0,14 | 61 | 8 | ~ 2-8 mp |

**A végleges Burst And Transient Source Experiment katalógus adatainak elemzése**

A Compton műhold 2000-ben befejezte működését. A Burst And Transient Source Experiment 2704 gammakitörést rögzített. A megfigyelt adatokat a Burst And Transient Source Experiment Végleges Katalógusban lehet elérni az interneten[16]. Az időtartam-táblázatban 1929 kitörésre található adat. 2002-ben ennek az 1929 kitörésnek az adatát vizsgáltam meg[17]. Ehhez az elemzéshez a maximum likelihood módszer használtam. Így minden adat önmagában szerepel az elemzés során.

A maximum likelihood módszer lényege, hogy az *a* paraméter valódi értékét azzal az *a*' értékkel becsüljük, mely, ha a paraméter valódi értéke volna, akkor éppen az adott minta bekövetkezése volna a legvalószínűbb az összes lehetséges minták között.

Ha adott az $x_1$, $x_2$, …. $x_n$ n-elemű mérési minta egy változóra, melynek sűrűségfüggvénye

$$f=f(x,a) \qquad (2)$$

és a mérési eredményekből az *a* értéket akarjuk becsülni, akkor az

$$l(x_1,x_2,\ldots x_n,a)=f(x_1,a)f(x_2,a)\ldots f(x_n,a) \qquad (3)$$

függvényt likelihood-függvénynek nevezzük.

A maximum likelihood módszer, hogy az *a* paraméter becslésére azt az értéket választjuk, ahol a likelihood-függvény értéke maximális. Az aszimptotikusan normális eloszlású becslések közül a maximum likelihood becslés a legjobb becslés[18].

A Burst And Transient Source Experiment végleges katalógus fent említett 1929 adatának az elemzésénél, az egy Gauss-görbe illesztés, hasonlóan az előzőekhez, nem adott egyezést. Két Gauss-görbét illesztve 5 illesztendő paraméter volt. Egy Gauss-görbét 3 paraméterrel írhatunk le; az eloszlás maximumhelye, szórása és súlya. Következésképpen két Gauss-görbe esetén 6 illesztendő paraméterünk van, de van egy feltétel is az amplitúdók összegére. Az amplitúdók összege vagy N (az észlelt objektumok száma), vagy ha a megtalálási valószínűség az illesztett függvény, akkor a teljes integrál értéke 1. Tehát az előbb említett feltétel akkor elégíthető ki, ha az illesztés során használt *súly* együtthatók összege 1.

Két Gauss-görbe illesztése esetén az L maximális értéke 12320,11, három Gauss-görbe illesztése esetén L maximális értéke 12326,25. A likelihood maximális értékének javulása várható, hiszen több paraméterrel írjuk le az illesztett elméleti görbét. Jelen esetben 3-mal nőtt a paraméterek száma. Ez esetben az L értékek különbségének a kétszerese egy 3 szabadsági fokú $\chi^2$-eloszlást követ[19]. Ezen

$$2(L_3-L_2)=\chi_3^2 \qquad (4)$$

képlet alapján számított valószínűség 0,5%.

A 2. táblázatban és a 3. táblázatban az illesztésekből kapott paraméterek találhatók. A szignifikanciát Monte-Carlo módszerrel ellenőriztem.

**2 táblázat.** A rövid és hosszú csoportok paraméterei.

| csoportok | középpont lgT90 | szórás | súly |
|---|---|---|---|
| rövid | -0,11 | 0,61 | 0,32 |
| hosszú | 1,54 | 0,43 | 0,68 |

[16] http://heasarc.gsfc.nasa.gov/docs/cgro/batse/BATSE_Ctlg/index.html (Letöltve: 2025. 03. 01.)
[17] Horváth 2002.
[18] Bronstein, I. N. and Szemengyajev i 1980. Paál 1992.
[19] Kendall, M. and Stuart 1976.

**3 táblázat.** A három csoport paraméterei.

| csoportok | középpont lgT90 | szórás | súly |
|---|---|---|---|
| rövid | -0,25 | 0,53 | 0,26 |
| közepes | 0,63 | 0,20 | 0,06 |
| hosszú | 1,55 | 0,42 | 0,68 |

Az 2. táblázat adataival, mint sűrűségfüggvénnyel, véletlenszerűen választottam 1929 számot. Ezen számokra elvégeztem a kétkomponens- és a háromkomponens-illesztést. Természetesen három komponens esetén a likelihood érteke magasabb lett. Ezt az eljárást elvégeztem még 99-szer (99x1929 véletlen számmal). A száz likelihood különbségeloszlását mutatja a 3. ábra.
Az eljárás során feltételeztem, hogy egy kétkomponensű mintából választok 1929 pontot. Feltételezve, hogy nincs harmadik komponens, a Monte-Carlo eljárással azt vizsgálom, hogy milyen eséllyel kaphatok a mérésnél tapasztalt likelihood javulást (6,14). Ez száz esetből egyszer történt meg, ami megerősíti a 3 szabadsági fokú $\chi^2$-eloszlást használva kapott 0,5% valószínűséget.
Hogy a harmadik populáció nem véletlen fluktuáció eredménye, azt az is alátámasztja, hogy a száz esetből szintén csak egyszer fordult elő, hogy a mérésnél kapott hat százaléknál nagyobb lett a csoport mérete. Az átlagos csoportméret a Monte-Carlo szimuláció során 2,5% volt.

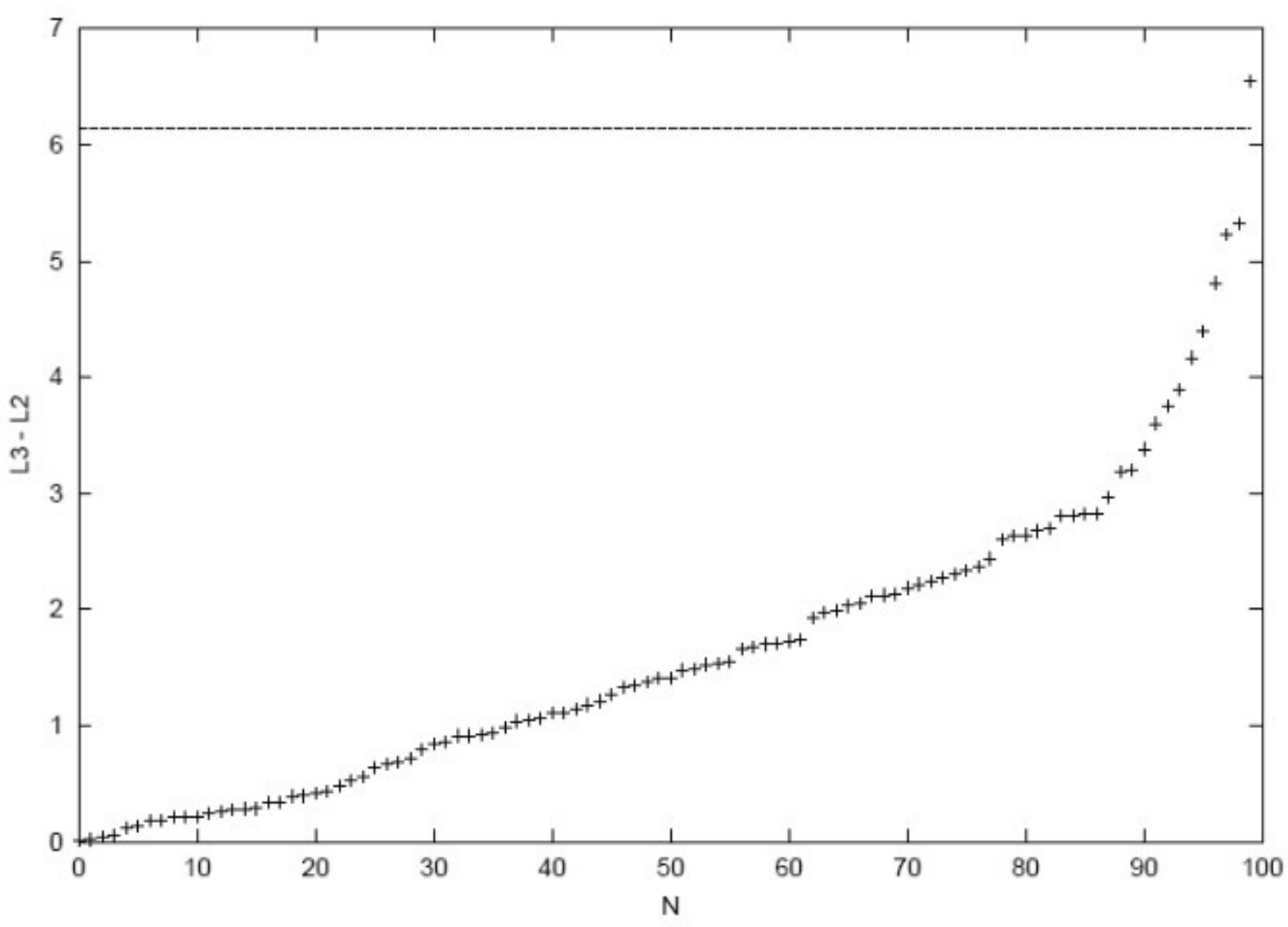

*3. ábra.* A Monte-Carlo szimulált likelihood különbségek növekvő sorrendbe rendezett eloszlása. A szaggatott vonallal jelzett 6,14-es érték az elemzésben kapott likelihood javulás. A Monte-Carlo szimulációkból ezt az értéket csak egy haladta meg.
*Forrás:* HORVÁTH 2002

**Vizsgálatok a T50 adatok felhasználásával**

Először a teljes Burst And Transient Source Experiment katalógus 1929 kitörésének a T90 adatán ismételtem meg a számításokat maximum likelihood módszerrel. Az eredmény teljes mértékben megegyezett az irodalomban közöltekkel. Két csoportot feltételezve a likelihood maximuma 12320,11, három Gauss-komponenst feltételezve 12326,25 volt. A különbség 6,14, mely 99,5%-os szignifikanciának felel meg. A legjobban illeszkedő paraméterek is azonosak voltak az irodalomban előzőleg közöltekkel.

Az 1929 darab T50 adattal is elvégeztem a számításokat. Két csoportot feltételezve a likelihood maximuma 12216,14, három Gauss-komponenst feltételezve 12221,61 volt. A különbség 5,47, mely 98,8%-os szignifikanciának felel meg.
1998-ban a Burst And Transient Source Experiment 3B katalógus 797 kitörésének T90 eloszlásában mutattam ki a harmadik komponenst 99,98% szignifikanciával[20]. Itt jegyzem meg, hogy ugyanezen 797 gammakitörés adatait használva egy másik csoport ezzel egy időben mutatta ki a harmadik komponens létét[21]. Tizennégy oldalas, részletes cikkükben a két keménységi hányados, a csúcsfényesség és a teljes fluxus mellett nem csak T90-t, hanem T50-et is használták, nem külön-külön, hanem egyszerre mind a hat változót. E hatdimenziós térben három csoportot találtak. Eredményük időtartamtengelyre vett vetülete nagyfokú egyezést mutatott az általam előzőleg közölt eredménnyel.

A 797 gammakitörés T90 és T50 adatainak eloszlásának illesztését most újra elvégeztem. Az illesztést maximum likelihood módszerrel végeztem. T90 esetén két csoportot feltételezve a likelihood maximuma 4364,52, három Gauss-komponenst feltételezve 4375,12 volt. A különbség 10,6, mely 99,99%-os szignifikanciának felel meg. A csoportparamétereket a 4. táblázat tartalmazza.

**4 táblázat.** Három komponens és T90 esetén a kapott csoport paraméterek. $q_l$ a csoport súlyaránya.

| csoport | $q_l$ | *log T90* | *szórás* |
|---|---|---|---|
| rövid | 0,28 | -0,35 | 0,49 |
| közepes | 0,07 | 0,61 | 0,14 |
| hosszú | 0,65 | 1,53 | 0,40 |

T50 esetén két csoportot feltételezve a likelihood maximuma 4341,6, három Gauss-komponenst feltételezve 4354,5 volt. A különbség 12,9, mely 99,999%-os szignifikanciának felel meg. Tehát a T50 értékek még nagyobb szignifikanciával mutatják a harmadik csoport lététt. A kapott csoportparamétereket az 5. táblázat tartalmazza.

**5 táblázat.** Három komponens és T50 esetén a csoport paraméterek. $q_l$ a csoport részaránya a megfigyelt összes gammakitörésből.

| csoport | $q_l$ | *log T50* | *szórás* |
|---|---|---|---|
| rövid | 0,27 | -0,76 | 0,45 |
| közepes | 0,05 | 0,22 | 0,09 |
| hosszú | 0,68 | 1,05 | 0,50 |

A két táblázatot összehasonlítva láthatjuk, hogy a két elemzés lényegében ugyanazt az eredményt adja. Ez által most kimutattam azt, hogy a Burst And Transient Source Experiment T50 adatok is igazolják a harmadik gammakitörés csoport létét.

**Összefoglalás**

A gammakitörések forrásainak a pontos ismeretét jól segíti osztályozásuk. Elterjedt nézet, hogy kétfajta forrása lehet a kitöréseknek. A hosszú kitörések a nagy tömegű gyorsan forgó csillagok végállapotával lehetnek kapcsolatban, a rövid kitörések a kompakt kettősök összeolvadása révén keletkezhetnek. Az ismert két típustól különböző gammakitörés fajtát többen is megpróbáltak azonosítani (például Pendleton et al., 1997.[22]). A közepesen hosszú lehetséges gammakitörés-csoportnak kiterjedt irodalma van.[23]

[20] HORVÁTH 1998.
[21] MUKHERJEE 1998.

Korábbi munkáimban megvizsgáltam a Burst And Transient Source Experiment 3B katalógus 797 gammakitörés időtartamainak eloszlását. Megmutattam, hogy a háromkomponensű eloszlás szignifikáns javulás a kétkomponensűhöz képest. Majd elemeztem a végleges Burst And Transient Source Experiment katalógust is. A katalógusban szereplő 1929 gammakitörés időtartameloszlását megvizsgáltam és megmutattam, hogy a háromkomponensű eloszlás szignifikáns javulás a kétkomponensűhöz képest.

**Jelen cikkemben mind a 3B mind a végleges Burst And Transient Source Experiment katalógus T50 adatait elemeztem. Hasonlóan a T90-nel kapcsolatos vizsgálatokhoz itt is sikerült kimutatnom, hogy a T50 adatok is igazolják a harmadik gammakitörés csoport létezését.**

**Internetes források**

---

[22] PENDLETON 1997.

[23] BALASTEGUI, A. – RUIZ-LAPUENTE, P. AND CANAL 2001. HORVÁTH 2002. MÉSZÁROS 2006. HORVÁTH 2009. ZITOUNI 2015. TARNOPOLSKI 2016. HORVÁTH 2020.